\documentclass[showpacs,floatfix]{revtex4}

\usepackage{amsmath}
\usepackage{graphicx}

\newcommand{\be}{\begin{equation}}
\newcommand{\ee}{\end{equation}}
\newcommand{\bse}{\begin{subequations}}
\newcommand{\ese}{\end{subequations}}

\newcommand{\To}{T_{\text{o}}}

\newcommand{\btheta}{\boldsymbol{\theta}}
\newcommand{\bxi}{\boldsymbol{\xi}}
\newcommand{\av}[1]{{\left\langle#1\right\rangle}}
\newcommand{\md}{\mathrm{d}}

\newcommand{\F}{{\mathcal F}}
\newcommand{\Fo}{{\mathcal F}_0}
\newcommand{\G}{{\mathcal G}}
\newcommand{\Go}{{\mathcal G}_0}
\newcommand{\Ho}{{\mathcal H}_0}

\newcommand{\hchc}{\av{h_c^2}}
\newcommand{\hchs}{\av{h_c h_s}}
\newcommand{\hshs}{\av{h_s^2}}
\newcommand{\xhc}{\av{x h_c}}
\newcommand{\xhs}{\av{x h_s}}

\newcommand{\avaa}{\av{a^2}}
\newcommand{\avab}{\av{ab}}
\newcommand{\avbb}{\av{b^2}}

\begin{document}

\title{Searching for gravitational waves from known pulsars using the $\F$ and $\G$ statistics}

\author{Piotr Jaranowski}

\address{Faculty of Physics, University of Bia{\l}ystok,
Lipowa 41, 15-424 Bia{\l}ystok, Poland}

\author{Andrzej Kr\'olak}

\address{Institute of Mathematics, Polish Academy of Sciences, \'Sniadeckich 8,
00-956 Warsaw, Poland}

\begin{abstract}

In searches for gravitational waves emitted by known isolated pulsars in data collected by a detector
one can assume that the frequency of the wave, its spindown parameters,
and the position of the source in the sky are known,
so the almost monochromatic gravitational-wave signal we are looking for
depends on at most four parameters:
overall amplitude, initial phase, polarization angle,
and inclination angle of the pulsar's rotation axis with respect to the line of sight.
We derive two statistics by means of which one can test
whether data contains such gravitational-wave signal:
the $\G$-statistic for signals which depend
on only two unknown parameters (overall amplitude and initial phase),
and the $\F$-statistic for signals depending on all four parameters.
We study, by means of the Fisher matrix, the theoretical accuracy
of the maximum-likelihood estimators of the signal's parameters
and we present the results of the Monte Carlo simulations we performed
to test the accuracy of these estimators.

\end{abstract}

\pacs{95.55.Ym, 04.80.Nn, 95.75.Pq, 97.60.Gb}

\maketitle

\section{Introduction}

We study the detection of almost monochromatic gravitational waves
emitted by known single pulsars in data collected by a detector.
Several such searches were already performed with data collected
by the LIGO and GEO600 detectors \cite{LSC04,LSC05,LSC07,LSC08,LSC10}.
We thus assume that the frequency of the wave (together with its
time derivatives, i.e.\ the spindown parameters)
and the position of the source in the sky are known.
The gravitational-wave signal we are looking for depends on at most
four (often called amplitude) parameters: overall amplitude, initial phase, polarization angle,
and inclination angle (of the pulsar's rotation axis with respect to the line of sight).

In Sec.\ 2 we introduce three statistics by means of which one can test
whether data contains a gravitational-wave signal: the $\mathcal{H}$-statistic
for completely known signals, the $\G$-statistic for signals which depend
on only two unknown parameters (overall amplitude and initial phase),
and the $\F$-statistic suitable for signals depending on all four amplitude parameters.
Both statistics $\G$ and $\F$ are derived from the maximum likelihood (ML) principle,
and the statistic $\G$ is independently obtained using Bayesian approach
and the composite hypothesis testing.
In Sec.\ 3 we study, by means of the Fisher matrix, the theoretical accuracy
of the ML estimators of the signal's parameters
and in Sec.\ 4 we present the results of the Monte Carlo simulations we performed
to test the accuracy of the ML estimators.

\section{Using the $\F$ and $\G$ statistics to perform targeted searches for gravitational waves from pulsars}
\label{sec:stat}

In the case when the signal $s(t)$ we are looking for is completely known,
the test that maximizes probability of detection
subject to a certain false alarm probability is the likelihood-ratio test,
i.e.\ we accept the hypothesis that the signal is present in detector's data $x$ if
\be
\label{lr1}
\Lambda(x) := \frac{p_1(x)}{p_0(x)} \geq \lambda_0,
\ee
where the likelihood function $\Lambda(x)$ is the ratio of probability densities
$p_1(x)$ and $p_0(x)$ of the data $x$ when the signal is respectively present or absent.
The parameter $\lambda_0$ is a threshold calculated from a chosen
false alarm probability. Assuming stationary and additive Gaussian noise with
one-sided spectral density constant (and equal to $S_0$) over the bandwidth of the signal,
the $\log$ likelihood function is approximately given by \cite{JKS98}
\be
\label{eq:LFpuln}
\ln\Lambda[x(t)] \cong 2 \frac{\To}{S_0}
\left( \av{x(t)s(t)} - \frac{1}{2} \av{s(t)^2} \right),
\ee
where $\To$ is the observation time and the time-averaging operator $\av{\cdot}$ is defined as
\be
\label{eq:tav}
\av{g} := \frac{1}{\To}\int^{\To}_{0}g(t)\,\md t.
\ee
Equation \eqref{eq:LFpuln} implies that the likelihood-ratio test \eqref{lr1}
can be replaced by the test
\be
{\mathcal H}[x(t)] := \av{x(t)s(t)} \geq \Ho,
\ee
where the optimal statistic ${\mathcal H}$ in this case is the {\em matched filter}
and $\Ho$ is the threshold for detection.

Suppose now that the signal $s(t;\btheta)$ depends on a set of unknown parameters $\btheta$,
then a suitable test can be obtained using a Bayesian approach
and {\em composite hypothesis} testing. The composite hypothesis in this case
is the hypothesis that when a signal is present it can assume any values of the parameters.
Assuming that the cost functions are independent of the values of the parameters,
we obtain the following Bayesian decision rule to choose the hypothesis
that the signal is present (see e.g.\ \cite{W71}, Chapter 5.9):
\be
\label{eq:BayesF}
\frac{1}{p_0(x)}\,{\int_\Theta}p_1(x;\btheta)\pi(\btheta)\,\md\btheta \geq \gamma_0,
\ee
where $\Theta$ is the parameter space on which $\btheta$ is defined
and $\pi(\btheta)$ is the joint a priori distribution of $\btheta$.
The expression on the left hand side of Eq.\ (\ref{eq:BayesF}) is know as
the {\em Bayes factor} and it is the ratio between the
posterior probability distribution on the signal parameters
marginalized over the parameters themselves (this is the signal model
{\em Bayesian evidence}) and the noise model which has no defining
parameters (this is the noise model {\em Bayesian evidence}).

As a template for the response of an interferometric detector
to the gravitational-wave signal from a rotating neutron star
we use the model derived in \cite{JKS98}.
This template depends on the set of following parameters:
$\btheta=(h_0,\phi_0,\psi,\iota,\mathbf{f},\delta,\alpha)$,
where $h_0$ is the dimensionless amplitude, $\phi_0$ is an initial phase,
$\psi$ is the polarization angle, $\iota$ is the inclination angle,
angles $\delta$ (declination) and $\alpha$ (right ascension) are equatorial coordinates
determining the position of the source in the sky,
and the `frequency vector' $\mathbf{f}:=(f_0,f_1,f_2,\dots)$
collects the frequency $f_0$ and the spindown parameters of the signal.
In the case of pulsars known from radio observations
we in general know the subset $\bxi=(\mathbf{f},\delta,\alpha)$ of the parameters $\btheta$.

Sometimes, like in the case of the Vela pulsar,
we also know from X-ray observations the values of the angles $\psi$ and $\iota$ (see \cite{NR04,NR08}
for observational results).
We then have only two unknown parameters: $h_0$ and $\phi_0$.
The response $s(t)$ of the detector to the gravitational wave
we can write in this case in the following form \cite{JKS98}:
\be
\label{eq:sig2}
s(t) = h_0 \cos\phi_0 \, h_c(t) + h_0 \sin\phi_0 \, h_s(t),
\ee
where $h_c$ and $ h_s$ are known functions of time,
\be
\begin{array}{l}
h_c(t) := A_+ \big(\cos2\psi\,h_1(t)+\sin2\psi\,h_2(t)\big)
- A_\times \big(\sin2\psi\,h_3(t)-\cos2\psi\,h_4(t)\big),
\\[1ex]
h_s(t) := -A_\times \big(\sin2\psi\,h_1(t)-\cos2\psi\,h_2(t)\big)
- A_+ \big(\cos2\psi\,h_3(t)+\sin2\psi\,h_4(t)\big).
\end{array}
\ee
Here the constants $A_+$ and $A_\times$ are
\be
\label{aa}
A_{+} := \frac{1}{2} (1 + \cos^2\iota),\quad
A_{\times} := \cos\iota,
\ee
and the four functions of time $h_k$ $(k=1,\ldots,4)$ depend only on parameters $\bxi$
and are defined as follows
\be
\label{eq:amps}
\begin{array}{ll}
h_1(t;\bxi) := a(t;\delta,\alpha) \cos \phi(t;\mathbf{f},\delta,\alpha),
&
h_2(t;\bxi) := b(t;\delta,\alpha) \cos \phi(t;\mathbf{f},\delta,\alpha),
\\[1ex]
h_3(t;\bxi) := a(t;\delta,\alpha) \sin \phi(t;\mathbf{f},\delta,\alpha),
&
h_4(t;\bxi) := b(t;\delta,\alpha) \sin \phi(t;\mathbf{f},\delta,\alpha),
\end{array}
\ee
where $a$, $b$ are the amplitude modulation functions and $\phi$ is
the phase modulation function. Their explicit forms are given in \cite{JKS98}.

Let us calculate the likelihood function for the signal (\ref{eq:sig2}).
Observing that the amplitude modulation functions $a$ and $b$
vary much more slowly than the phase $\phi$ of the signal
and assuming that the observation time is much longer
than the period of the signal we approximately have \cite{JKS98}
\be
\label{app1}
\begin{array}{l}
\av{h_1\,h_3} \cong \av{h_1\,h_4} \cong \av{h_2\,h_3} \cong \av{h_2\,h_4} \cong 0,
\\[1ex]
\av{h_1\,h_1} \cong \av{h_3\,h_3} \cong \frac{1}{2} A, \quad
\av{h_2\,h_2} \cong \av{h_4\,h_4} \cong \frac{1}{2} B, \quad
\av{h_1\,h_2} \cong \av{h_3\,h_4} \cong \frac{1}{2} C,
\end{array}
\ee
where we have introduced the time averages
\be
\label{ABCdef}
A := \av{a^2}, \quad
B := \av{b^2}, \quad
C := \av{ab}.
\ee
As a consequence of the above approximations we have the following
approximate expressions for the time averaged products of the functions
$h_c$ and $h_s$,
\be
\label{eq:simpl}
\av{h_c^2} \cong \av{h_s^2} \cong N,
\quad \av{h_c h_s} \cong 0,
\ee
where $N$ is a constant defined as
\begin{align}
N &:= \frac{1}{2} \Big(  A( A_+^2\cos^2 2\psi + A_{\times}^2\sin^2 2\psi)
+ B( A_+^2\sin^2 2\psi + A_{\times}^2\cos^2 2\psi)
\nonumber\\ &\qquad
+ C( A_+^2 - A_{\times}^2) \sin 4\psi \Big).
\end{align}
With the above approximations the likelihood function $\Lambda$ for the signal \eqref{eq:sig2}
can be written as
\be
\label{eq:LFpul}
\ln\Lambda[x(t);\phi_0,h_0] \cong 2 \frac{\To}{S_0}
\left( h_0 \cos\phi_0 \av{x(t)h_c(t)} + h_0 \sin\phi_0 \av{x(t)h_s(t)} - \frac{1}{2} h_0^2 N \right).
\ee
Let us also note that the optimal signal-to-noise ratio (SNR) $\rho$ for the signal \eqref{eq:sig2}
(see \cite{JKS98} for definition) can be approximately computed as
\be
\label{snr}
\rho \cong \sqrt{\frac{2\To}{S_0}\av{s(t)^2}} \cong \sqrt{\frac{2\To N}{S_0}} h_0.
\ee

It is natural to assume that the prior probability density of the phase parameter $\phi_0$
is uniform over the interval $[0,2\pi)$
and that it is independent of the distribution of the amplitude parameter $h_0$, i.e.
\be
\pi(\phi_0) = \frac{1}{2\pi}, \quad \phi \in [0,2\pi).
\ee
With the above assumptions the integral $\int_0^{2\pi} p_1(x;\phi_0,h_0)\pi(\phi_0)\md\phi_0$
can be explicitly calculated (see \cite{W71}, Chapter 7.2)
and we obtain the following decision criterion
\be
\label{eq:IGstat}
\exp\bigg(-\frac{h_0^2 N \To}{S_0}\bigg) \,
I_0\bigg(2 h_0 \sqrt{\frac{\To N}{S_0} \G[x(t)]}\bigg) \geq \gamma_0,
\ee
where $I_0$ is the modified Bessel function of zero order
and the statistic $\G$ is defined as
\be
\label{eq:Gstat}
\G[x(t)] := \frac{\To}{N S_0} \Big( \av{x(t)h_c(t)}^2 + \av{x(t)h_s(t)}^2 \Big).
\ee
The function on the left-hand side of Eq.\ (\ref{eq:IGstat}) is a monotonically increasing
function of $\G$ and it can be maximized if $\G$ is maximized independently
of the value of $h_0$. Thus the test
\be
\G[x(t)] \geq \Go,
\ee
provides a uniformly most powerful test with respect to the amplitude $h_0$.

When we have no a priori information about the parameters a standard method
is the {\em maximum likelihood} (ML) detection which consists
of maximizing the likelihood function $\Lambda[x(t);\btheta]$ with respect to
the parameters of the signal. If the maximum of $\Lambda$ exceeds a
certain threshold we say that the signal is detected.
The values of the parameters that maximize $\Lambda$
are said to be the ML estimators of the parameters of the signal.
For the case of signal (\ref{eq:sig2}) it is convenient to
introduce new parameters
\be
A_c := h_0 \cos\phi_0, \quad  A_s := h_0 \sin\phi_0.
\ee
Then one can find the ML estimators of the amplitudes
$A_c$ and $A_s$ in a closed analytic form,
\be
\hat{A}_{c} \cong \frac{\av{x h_c}}{N},
\quad
\hat{A}_{s} \cong \frac{\av{x h_s}}{N}.
\ee
It is easy to find that the estimators $\hat{A}_{c}$ and $\hat{A}_{s}$ are
unbiased and also that they are of minimum variance,
i.e.\ their variances attain the lower Cram\'er-Rao bound determined by the Fisher matrix.
The variances of both estimators are the same and equal to $1/N$.
Substituting the estimators $\hat{A}_{c}$ and $\hat{A}_{s}$ for the
parameters $A_c$ and $A_s$ in the likelihood function one
obtains a reduced likelihood function. This reduced likelihood function
is precisely equal to the $\G$-statistic given by Eq.\ (\ref{eq:Gstat}),
i.e.\ $\G[x(t)]=\ln\Lambda[x(t);\hat{A}_{c},\hat{A}_{s}]$.
The formula for the $\G$-statistic obtained without usage of the simplifying assumptions
\eqref{app1} is given in Appendix A.

When the all four parameters $(h_0,\phi_0,\psi,\iota)$ are unknown one can
introduce new parameters $A_k$ ($k = 1,\ldots,4$) that are functions
of $(h_0,\phi_0,\psi,\iota)$ such that the response $s(t)$ takes the form
\be
\label{eq:sig}
s(t) = A_1 \, h_1(t) + A_2 \, h_2(t) + A_3 \, h_3(t) + A_4 \, h_4(t),
\ee
where the functions $h_k$ are given by Eqs.\ (\ref{eq:amps})
and the parameters $A_k$ read
\be
\label{eq:ampone}
\begin{array}{l}
A_1 := h_{0+}\cos2\psi\cos\phi_0 - h_{0\times}\sin2\psi\sin\phi_0,
\\[1ex]
A_2 := h_{0+}\sin2\psi\cos\phi_0 + h_{0\times}\cos2\psi\sin\phi_0,
\\[1ex]
A_3 := -h_{0+}\cos2\psi\sin\phi_0 - h_{0\times}\sin2\psi\cos\phi_0,
\\[1ex]
A_4 := -h_{0+}\sin2\psi\sin\phi_0 + h_{0\times}\cos2\psi\cos\phi_0;
\end{array}
\ee
here $h_{0+}:=h_0\,A_+$ and $ h_{0\times}:=h_0\,A_\times$ [see Eq.\ \eqref{aa}].
The ML estimators of $A_k$ can again be obtained in an explicit analytic form
and the reduced likelihood  function is the $\F$-statistic given by (see \cite{JKS98} for details)
\begin{align}
\label{eq:Fstat}
\F[x(t)] := \ln\Lambda[x(t);\hat{A}_1,\ldots,\hat{A}_4] \cong &\,\frac{2\To}{S_0 D}
\Big( B\, (\av{x h_1}^2 + \av{x h_3}^2) + A\, (\av{x h_2}^2 + \av{x h_4}^2)
\nonumber\\&\qquad\quad
- 2C\, (\av{x h_1} \av{x h_2} + \av{x h_3} \av{x h_4}) \Big),
\end{align}
where $D:=AB-C^2$. The test
\be
\F[x(t)] \geq \Fo
\ee
is not a uniformly most powerful test with respect to unknown parameters $(h_0,\phi_0,\psi,\iota)$.
It was recently shown that uniform a priori distributions of $(h_0,\phi_0,\psi,\cos\iota)$
lead to a statistic that can be more powerful than $\F$ \cite{PK09}.

\begin{figure}
\begin{center}
\scalebox{0.5}{\includegraphics{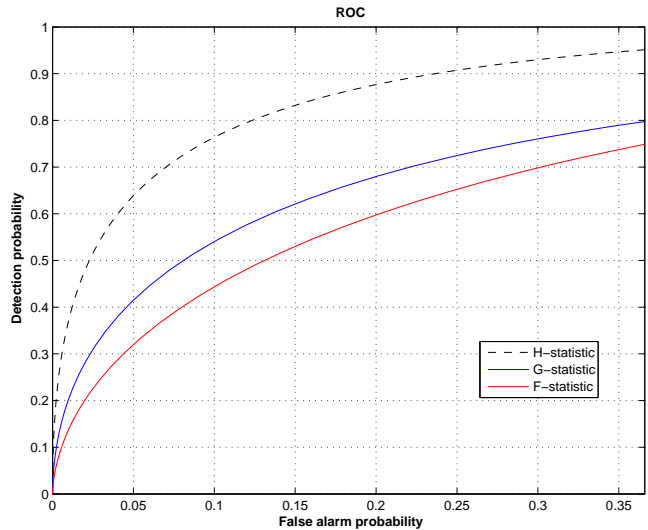}}
\caption{\label{fig:roc}
Receiver operating characteristic (ROC) for the statistics ${\mathcal H}$, $\G$, and $\F$
for the optimal signal-to-noise ratio $\rho=2$.}
\end{center}
\end{figure}

In Fig.\ \ref{fig:roc} we have plotted the receiver operating characteristics (ROC)
for the three statistics ${\mathcal H}$, $\G$, and $\F$ considered in the present section.

\section{The Fisher matrix}

Using the Fisher matrix we can assess the accuracy of the parameter estimators.
We have two theorems that can loosely be stated as follows.

\newtheorem{theorem}{Theorem}

\begin{theorem}[Cram\`er-Rao bound]
The diagonal elements of the inverse of the Fisher matrix
are lower bounds on the variances of unbiased estimators of the parameters.
\end{theorem}

\begin{theorem}
Asymptotically (i.e.\ when the SNR tends to infinity)
the ML estimators are unbiased and their covariance matrix is equal to the inverse of the Fisher matrix.
\end{theorem}

For an almost monochromatic signal $s=s(t;\btheta)$,
which depends on the parameters $\btheta=(\theta_1,\ldots,\btheta_m)$,
the elements of the Fisher matrix $\Gamma$ can be approximately calculated
from the formula
\be
\label{mFisher}
\Gamma_{{\theta_i}{\theta_j}} \cong \frac{2\To}{S_0}
\av{\frac{\partial s}{\partial\theta_i}\frac{\partial s}{\partial\theta_j}},
\quad i,j=1,\ldots,m.
\ee

In the case when only the parameters $h_0$ and $\phi_0$ are unknown ($\G$-statistic search),
the Fisher matrix can be computed easily from Eqs. \eqref{eq:sig2} and \eqref{mFisher}.
It is diagonal and the standard deviations of the parameters
defined as the square roots of the diagonal elements of the inverse of the Fisher matrix read:
\be
\frac{\sigma_{h_0}}{h_0} = \frac{1}{\rho}, \quad
\sigma_{\phi_0} = \frac{1}{\rho},
\ee
where $\rho$ is the optimal SNR [given in Eq.\ \eqref{snr}].

\begin{figure}
\begin{center}
\scalebox{0.75}{\includegraphics{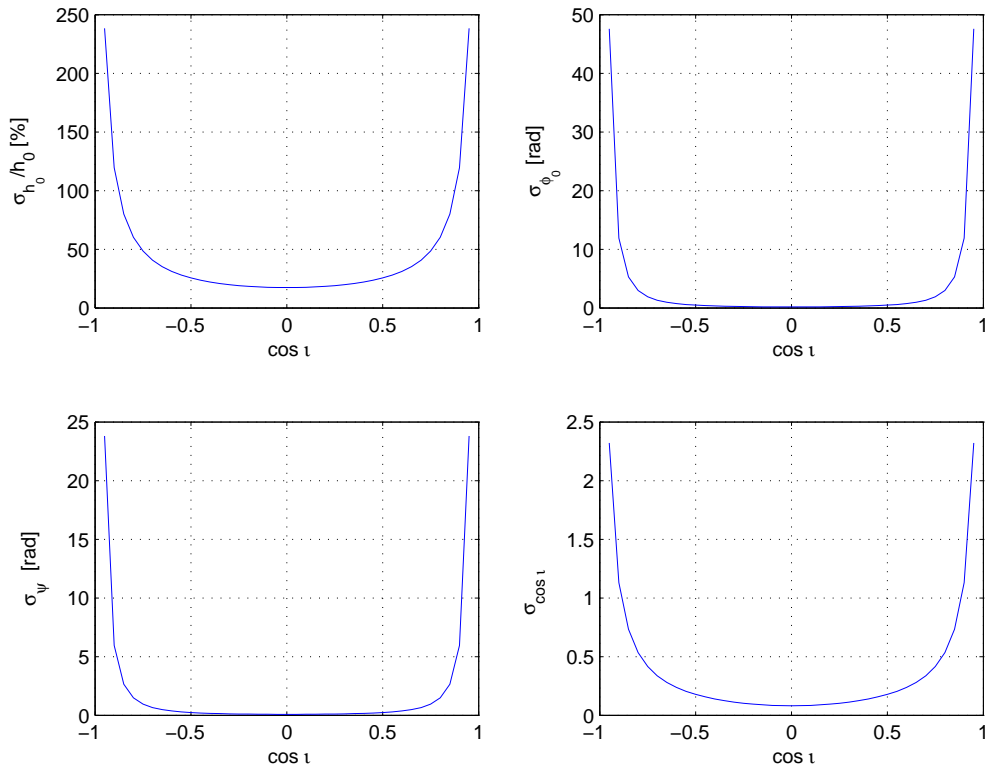}}
\caption{\label{fig:pulsar2}
Dependence of standard deviations (calculated from the Fisher matrix)
of the parameters $h_0$, $\phi_0$, $\psi$, and $\cos\iota$
on the cosine of the inclination angle $\iota$.
We have taken $\phi_0=4.03$ and $\psi=-0.22$
(values of other parameters needed to perform the computation of the Fisher matrix
are listed in the text of Sec.\ 3).}
\end{center}
\end{figure}

When all the four amplitude parameters $h_0$, $\phi_0$, $\psi$, and $\iota$ are unknown
($\F$-statistic search), the Fisher matrix can be computed by means of formulas given in Appendix B.
In this case it is not diagonal, indicating that the amplitude parameters are correlated.
The quantities $\sigma_{h_0}/h_0$, $\sigma_{\phi_0}$, $\sigma_{\psi}$, $\sigma_{\iota}$
(where the standard deviations again are defined as square roots of diagonal elements of the inverse of the Fisher matrix)
have rather complicated analytical form but they possess a number of simple properties.
They are inversely proportional to the overall amplitude $h_0$,
independent on the initial phase $\phi_0$, and very weakly dependent on $\psi$,
however there is a strong dependence on $\iota$.

In Fig.\ \ref{fig:pulsar2} we have shown the dependence of the standard deviations
on the cosine of the inclination angle $\iota$.
The time averages from Eqs.\ \eqref{ABCdef} (needed to compute the Fisher matrix)
were computed here for the location of the Virgo detector \cite{Virgo08}
and for a randomly chosen position of the source in the sky.
We have also taken $h_0=6.0948\times10^{-2}$, $\To=441610$~s, and $S_0=2$~Hz$^{-1}$,
which corresponds to the SNR $\rho\cong28.64\sqrt{2N}$ [see Eq.\ \eqref{snr}].
The same time averages and the values of $\To$, $h_0$, $S_0$ were used
in the Monte Carlo simulations described in Sec.\ 4.
We see in Fig.\ \ref{fig:pulsar2} that the standard deviations become singular when $\cos\iota=\pm1$.
This singularity originates from the degeneracy of the amplitude parameters for $\cos\iota=\pm1$.
In this case the amplitude parameters from Eqs.\ \eqref{eq:ampone} become
\be
\label{eq:ampone1}
A_{1} = h_0 \cos(2\psi\pm\phi_0), \quad
A_{2} = h_0 \sin(2\psi\pm\phi_0), \quad
A_{3} = \mp A_{2}, \quad
A_{4} = \pm A_{1}.
\ee
Thus only two of them are independent. Therefore the determinant of
the 4-dimensional Fisher matrix is equal to zero at $\cos\iota=\pm1$
and consequently its inverse does not exist in this case.

\section{Monte Carlo simulations}

We have performed two Monte Carlo simulations in order to test the performance of the ML estimators.
We have compared the simulated standard deviations of the estimators with the ones obtained from the Fisher matrix.
In particular we have investigated the behavior of the ML estimators near the Fisher matrix singularity at $\cos\iota=\pm1$.
In each simulation run we have generated the signal using Eq.\ \eqref{eq:sig},
we have added it to a white Gaussian noise,
and we have estimated the amplitude parameters using the $\F$-statistic.
Each simulation run was repeated 1000 times for different realizations of the noise.

\begin{figure}
\begin{center}
\scalebox{0.75}{\includegraphics{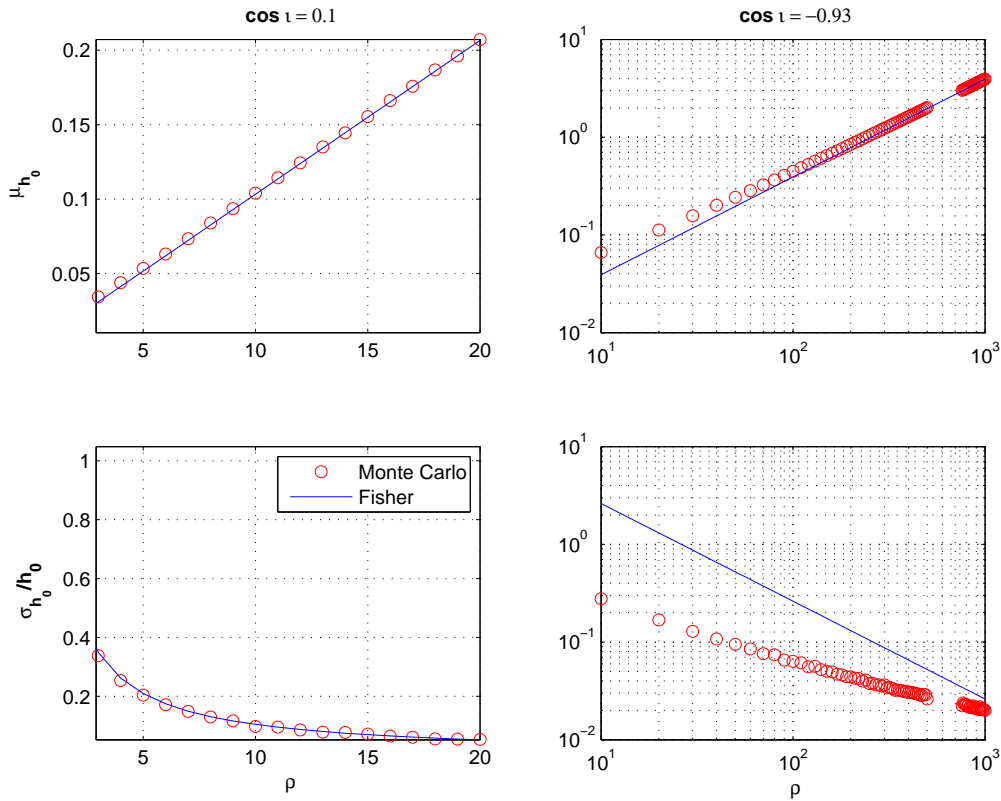}}
\caption{\label{fig:pulsar2snr}
Mean and normalized standard deviation of the ML estimator of the amplitude $h_0$ as a function of the SNR.
The top two panels are the means of the estimator
for the two values of $\cos\iota$. The continuous line is the true value and the
circles are results of the simulation for 1000 realizations of the noise.
The bottom two panels are the standard deviations. The continuous line is
obtained form the Fisher matrix whereas the circles are results of the simulation.
We have taken $\phi_0=4.03$ and $\psi=-0.22$.}
\end{center}
\end{figure}

In the first simulation we have investigated the bias and the standard deviation
of the ML estimator of the amplitude parameter $h_0$ as functions of the SNR
for the two cases: $\cos\iota=0.1$ and $\cos\iota=-0.93$.
The results are presented in Fig.\ \ref{fig:pulsar2snr}.
For the first case the ML estimator is nearly unbiased
and its standard deviation is close to the one predicted by the Fisher matrix even for low SNRs.
In the second case the simulation shows considerable bias of the estimator
and its standard deviation lower than the one predicted by the Fisher matrix.
However, Theorem 2 is satisfied in the
second case. For $\cos\iota$ close to $\pm 1$ we have to go to SNR
$\sim 1000$ in order for the ML estimator to be unbiased
and its standard deviation close to the one given by the Fisher matrix.

\begin{figure}
\begin{center}
\scalebox{0.75}{\includegraphics{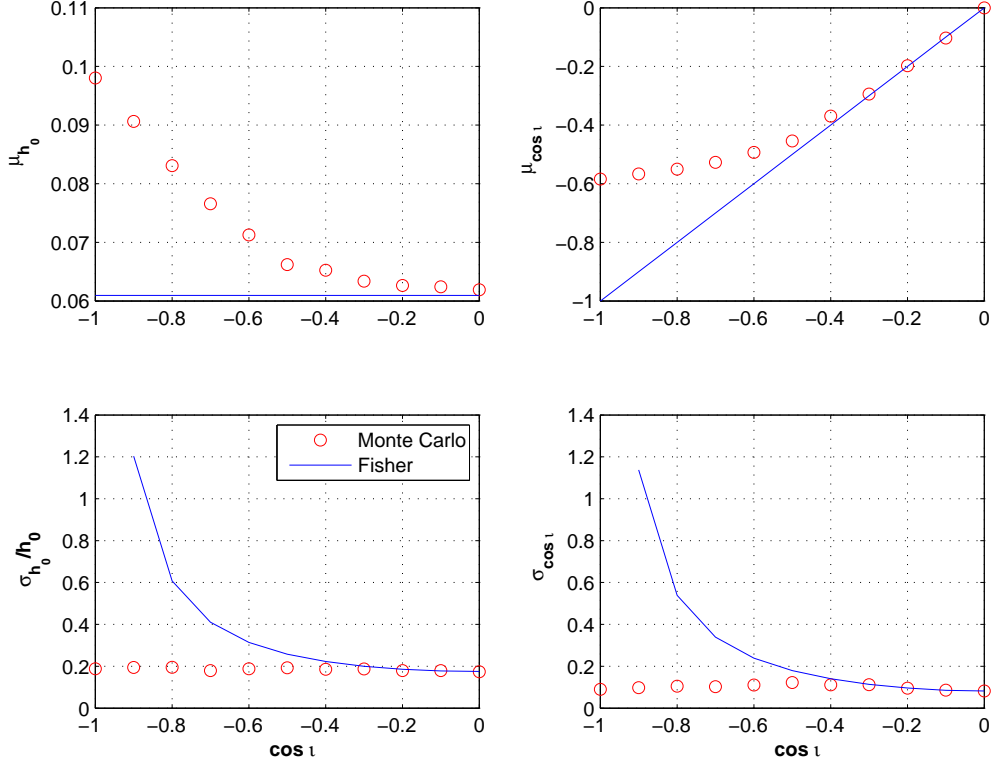}}
\caption{\label{fig:pulsar2cin}
Means and normalized standard deviations of the ML estimators of $h_0$ and $\cos\iota$
as functions of $\cos\iota$.
The top two panels are the means of the estimators.
The continuous lines are the true values
and the circles are results of the simulation for 1000 realizations of the noise.
The bottom two panels are the standard deviations. The continuous lines are
obtained form the Fisher matrix whereas the circles are results of the simulation.
We have assumed $\phi_0=4.03$, $\psi=-0.22$, and $\rho=15.6$.
Plots for $0\le\cos\iota\le+1$ (not shown here)
are mirror images of the plots for $-1\le\cos\iota\le0$.}
\end{center}
\end{figure}

In the second simulation, illustrated in Fig.\ 4,
we have investigated the bias and the standard deviation
of the ML estimators of the amplitude parameters $h_0$ and $\cos\iota$ as functions of $\cos\iota$
for the fixed SNR $\rho=15.6$.
We find that for $|\cos\iota|<0.5$ the biases are less than 10\%
and the Fisher matrix overestimates the standard deviations also by less than 10\%.
We see that over the whole range of $\cos\iota$  the standard deviations of the parameters are roughly constant
whereas the biases increases as the $|\cos\iota|$ increases.
At $\cos\iota\pm1$ the amplitude $h_0$ is overestimated by almost a factor of 2.

One reason why Theorem 1 does not apply here is that it holds for unbiased
estimators. Also a more precise statement of Theorem 1
(see e.g.\ Theorem 8 in \cite{JKbook}) requires that the Fisher matrix $\Gamma$
is positive definite for all values of parameters. This last assumption is clearly
not satisfied here as $\det\Gamma=0$ for $\cos\iota=\pm1$.

\appendix

\section{The general form of the $\G$-statistic}

It is not difficult to obtain the $\G$-statistic without simplifying assumptions \eqref{app1}.
The estimators of the amplitude parameters $A_c$ and $A_s$ are then given by
\be
\hat{A}_{c} = \frac{\hshs \xhc - \hchs \xhs}{\hchc \hshs - \hchs^2},
\quad
\hat{A}_{s} = \frac{\hchc \xhs - \hchs \xhc}{\hchc \hshs - \hchs^2},
\ee
and the general form of the $\G$-statistic reads
\be
\G[x(t)] \cong \frac{\To}{S_0}
\left( \frac{\hshs \xhc^2 - 2 \hchs \xhc \xhs + \hchc \xhs^2}{\hchc \hshs - \hchs^2} \right).
\ee

\section{Fisher matrix for amplitude parameters}

Let us consider the gravitational-wave signal $s$ of the form
\be
\label{sig}
s(t;\mathbf{A}) = \sum^4_{k=1} A_k\,h_k(t),
\ee
where the vector $\mathbf{A}$ collects the amplitude parameters,
$\mathbf{A}:=(A_1,A_2,A_3,A_4)$,
and the \emph{known} functions $h_k$ ($k=1,\dots,4$) are given in Eqs.\ \eqref{eq:amps}.
We further assume, as in Sec.\ 2, that the noise spectral density is constant (and equal to $S_0$)
over the bandwidth of the signal and that the approximations \eqref{app1} are valid.
Then the Fisher matrix for the signal's parameters $\mathbf{A}$ reads
\be
\Gamma(\mathbf{A}) \cong \frac{\To}{S_0}
\begin{pmatrix}
\avaa & \avab & 0 & 0 \\[1ex]
\avab & \avbb & 0 & 0 \\[1ex]
0 & 0 & \avaa & \avab \\[1ex]
0 & 0 & \avab & \avbb
\end{pmatrix},
\ee
and its inverse is qual to
\be
\Gamma(\mathbf{A})^{-1} \cong \frac{S_0}{\To\big(\avaa\avbb-\avab^2\big)}
\begin{pmatrix}
\avbb & -\avab & 0 & 0 \\[1ex]
-\avab & \avaa & 0 & 0 \\[1ex]
0 & 0 & \avbb & -\avab \\[1ex]
0 & 0 & -\avab & \avaa
\end{pmatrix}.
\ee
Let us introduce new set of parameters $\btheta:=(h_0,\phi_0,\psi,\iota)$.
Then the Fisher matrix $\Gamma(\btheta)$ for these parameters can be computed as
($\mathsf{T}$ denotes here matrix transposition)
\be
\Gamma(\btheta) = J^\mathsf{T} \cdot \Gamma(\mathbf{A})\cdot J,
\ee
where the Jacobi $4\times4$ matrix $J$ has elements $\partial A_i/\partial\theta_j$
($i,j,=1,\ldots,4$), which can be computed by means of Eqs. \eqref{eq:ampone}.

\section*{Acknowledgments}

This work was supported by the MNiSW grant no.\ N N203 387237.
A.K.\ would like to acknowledge hospitality
of the Max Planck Institute for Gravitational Physics in Hannover,
Germany, where part of this work was done.
We would like to thank members of the LSC-Virgo CW data analysis group for helpful discussions.


\newcommand{\PR}{Phys.\ Rev.\ }
\newcommand{\PRL}{Phys.\ Rev.\ Lett.\ }
\newcommand{\CQG}{Class.\ Quantum Grav.\ }

\end{document}